# Arginine-rich peptides destabilize the plasma membrane, consistent with a pore formation translocation mechanism of cell penetrating peptides

H. D. Herce[1,2,4], A. E. Garcia[1,2§], J. Litt[2,3], R. S. Kane[2,3], P. Martin[4], N. Enrique[4], A. Rebolledo[4], and V. Milesi[4].

[1]Department of Physics, Applied Physics and Astronomy, [2]Center for Biotechnology and Interdisciplinary Studies, and [3]Department of Chemical and Biological Engineering, Rensselaer Polytechnic Institute, Troy, NY, USA. 12180
[4]Universidad Nacional de La Plata, Consejo Nacional de Investigaciones Científicas y Técnicas, La Plata, Argentina.

[§]Corresponding author: angel@rpi.edu

**Abstract**

Recent molecular dynamics simulations (Herce and Garcia, PNAS, **104**: 20805 (2007)) have suggested that the arginine-rich HIV Tat peptides might be able to translocate by destabilizing and inducing transient pores in phospholipid bilayers. In this pathway for peptide translocation, arginine residues play a fundamental role not only in the binding of the peptide to the surface of the membrane but also in the destabilization and nucleation of transient pores across the bilayer, despite being charged and highly hydrophilic. Here we present a molecular dynamics simulation of a peptide composed of nine arginines (Arg-9) that shows that this peptide follows the same translocation pathway previously found for the Tat peptide. We test this hypothesis experimentally by measuring ionic currents across phospholipid bilayers and cell membranes through the pores induced by Arg-9 peptides. We find that Arg-9 peptides, in the presence of an electrostatic potential gradient, induce ionic currents across planar phospholipid bilayers, as well as in cultured osteosarcoma cells and human smooth muscle cells freshly isolated from the umbilical artery. Our results suggest that the mechanism of action of Arg-9 peptide involves the creation of transient pores in lipid bilayers and cell membranes.

## Introduction

Cell penetrating peptides (CPPs) are short sequences of amino acids (<30) capable of entering most mammalian cells. CPPs have the special property of carrying with them cargoes of a wide range of molecular size such as proteins, oligonucleotides, and even 200 nm liposomes (1-8). Many CPPs are highly cationic and hydrophilic, exhibiting no or relatively low amphipathicity when compared to other peptides that are known to interact with and permeabilize phospholipid membranes.

The translocation mechanism by which these peptides are able to enter cells has remained elusive since their discovery (2, 3, 9, 10). There is ample evidence suggesting that the uptake is independent of metabolic energy and it does not involve any specific cell receptor (1, 3). Other reports indicate that the uptake may involve lipid raft mediated endocytotic pathways (11-13). Even when the uptake could initially follow an endocytotic pathway, arg-rich CPPs are still able to breach the vesicle membrane barrier to reach for example the cell nucleus (14).

A common feature of cationic CPPs is that they form a rapid and tight interaction with extracellular glycosaminoglycans, such as heparan sulfate, heparin, and chondroitin sulfate B (14). However, whereas a significant part of the uptake of CPPs might involve heparan sulfate receptors, efficient internalization is observed even in their absence (15). Furthermore, these peptides are capable of entering giant unilamellar vesicles (GUVs) composed of model phospholipid membranes (16-19). These results suggest that these peptides can directly interact with the phospholipid bilayer altering its resting structure and that these changes facilitate peptide translocation. Taken together this evidence suggests that the binding of highly cationic CPPs to anionic plasma membrane components, such as phosphate and sulphate groups, could be an important step for effective peptide translocation across a phospholipid bilayer or the cell membrane.

The relevance of peptide-phosphate interactions can be understood through a theoretical model that we recently proposed for the translocation of the Tat peptide (20). This model shows how these peptides may be able to nucleate a pore and passively diffuse across a phospholipid bilayer. This model shows the special importance of the arginine amino acids in the translocation mechanism. According to this model, the arginine and lysine amino acids initially bind to the phospholipid phosphate groups producing strong distortions to the bilayer relative to their resting structure. This is followed by the translocation of a single arginine amino acid towards the distal side attracted by phosphate groups and this event nucleates a water pore. The relevance of the Arg-phosphate interaction has been highlighted in several reports that indicate for example that polyarginine peptides translocate more efficiently than polylysine peptides and the Tat peptide (10, 21). The interaction of arginine amino acids with phosphate groups of the plasma membrane plays a fundamental role in other physiological process such as the voltage gating of potassium ion channels (22, 23). It has even been suggested that the use of arginine in voltage sensors might be an adaptation to the phospholipid composition of cell membranes (24).

Several arginine rich antimicrobial peptides can also form pores in lipid bilayers and bacterial cell membranes (25, 26). It is known that antimicrobial peptides are able to translocate across the bacterial plasma membrane in an energy-independent manner. Since CPPs also seem capable of translocation in an energy independent manner, they could share a similar pore opening mechanism (1, 3, 20, 27). However, an important difference between hydrophilic CPPs and antimicrobial peptides is the lack of hydrophobic residues in the former case. It is known that the phospholipid bilayer of the cell membrane is not permeable to most ions and hydrophilic molecules. Therefore, it is generally assumed that a substantial hydrophobic content is necessary to directly interact with the core of the plasma membrane and nucleate a pore. Although CPPs are extremely hydrophilic, the model proposed for CPPs (20) suggests that arginine amino acids can efficiently disrupt and cross a phospholipid bilayer and nucleate a pore even in the absence of hydrophobic amino acids.

It has been proven that polyarginine peptides composed of nine arginine amino acids (Arg-9) are able to translocate across cells very efficiently (28). This raises the question if molecular dynamics simulations would show a similar pathway for this highly hydrophilic and charged peptide as predicted for the Tat peptide. The simplicity of Arg-9 over other CPPs also helps isolate the role of the Arg sidechain on the translocation without the effect of other sidechains. Therefore, in this work we study computationally the translocation of Arg-9 through a lipid bilayer and show that the translocation mechanism is the same as for the Tat peptide.

Based on the theoretical translocation pathway found for Tat and Arg-9 the central hypothesis of the experimental work presented here is the following: *If arginine-rich CPPs disturb and induce pores in phospholipid membranes then ions should be able to flow across lipid membranes through these pores. Therefore, applying an electrostatic potential should produce ionic currents through the pores induced by CPPs across phospholipid bilayers and cell membranes.* Consistent with this hypothesis, we detected ionic currents induced by Arg-9 on (a) model phospholipid membranes using the method of *planar phospholipid bilayers* (29), and (b) on freshly isolated human umbilical artery (HUA) smooth muscle cells and on cultured osteosarcoma cells using the *patch clamp technique* (30). We found that the ionic permeability of phospholipid bilayers and cell membranes is increased by the presence of the Arg-9 peptides. We also explored other peptides and solution conditions that alter this permeability.

This work covers systems with different degrees of complexity, ranging from simulations of very simple systems to experiments on live mammalian cells. The molecular dynamics studies reported previously for the Tat peptides (20) and here for Arg-9 peptides were conducted on simple systems containing a single phospholipid composition (DOPC), a few peptides and water. Cell membranes and experimental model membranes are usually composed of multiple lipids (27, 31, 32), cell receptors and modified lipids. Here, we study experimentally a series of systems that are as simple as the systems modeled computationally (e.g., a DOPC planar phospholipid bilayer); systems that contain lipid mixtures (e.g., 3:1 DOPC: DOPG); and mammalian cells under different salt and pH conditions. This allow us to present physical, chemical and biological details of the interactions between Arg-rich peptides and the cell membrane that we believe are essential to unveil the cell translocation pathway for CPPs.

## Methods
*Molecular dynamics simulations*

Molecular dynamics simulations were performed to study the translocation mechanism of Arg-9 peptides across model membranes. The Arg-9 peptides were placed in a periodically repeating box containing a pre-equilibrated lipid membrane composed of 1,2-Dioleoyl-*sn*-Glycero-3-Phosphocholine (DOPC) lipids, and water molecules (33, 34). The peptides were placed near one side of the bilayer such that the Arg-9 peptides would bind mostly to one layer of the bilayer. These configurations are away from equilibrium. Here, we explored how the systems relax from these configurations. Due to periodic boundary conditions, there are two paths by which an Arg-9 peptide on one layer can move to bind the other layer: one which requires translocation, and another which requires diffusion of the Arg-9 peptide from the initial configuration near the water-bilayer boundary on one layer to the other layer. Our calculations include a large number of water molecules such that binding to the proximal layer, given the initial conditions that we selected, is favored. The total length of the simulation was 500ns.

The simulations were performed using the GROMACS package (35), on a cluster of dual Opteron processors. The simulated system consists of 4 Arg-9 peptides, 92 DOPC phospholipid molecules and 8795 water molecules. The overall temperature of the water, lipids, and peptides were kept constant, coupling independently each group of molecules at 323 K with a Berendsen thermostat (36). The pressure was coupled to a Berendsen barostat at 1 atm separately in every dimension (36). The temperature and pressure time constants of the coupling were 0.2 ps and 2 ps, respectively. An external electric field of 0.05 V nm$^{-1}$ was included pointing towards the distal side of the bilayer to qualitatively take into account the external electrostatic potential included in the experiments. The integration of the equations of motion was performed using a leap frog algorithm with a time step of 2 fs. Periodic boundary conditions were implemented in all systems. A cut-off of 1 nm was implemented for the Lennard-Jones and the direct space part of the Ewald sum for Coulombic interactions. The Fourier space part of the Ewald splitting was computed using the particle-mesh Ewald method (37), with a grid length of 0.11 nm on the side and a cubic spline interpolation. Periodic boundary conditions and Ewald summations that do not include the surface term ensure system electro neutrality (38). We used the SPC/E (39), the lipid parameters were from Berger et al (33), and the peptide parameters were from the GROMACS force field (35).

*Planar lipid bilayers*
1,2-Dioleoyl-*sn*-Glycero-3-Phosphocholine (DOPC), and 1,2-Dioleoyl-*sn*-Glycero-3-[Phospho-*rac*-(1-glycerol)] (DOPG), were purchased from Avanti Polar Lipids, Tat peptide, penetratin and Arg-9 were purchased from Global Peptides, Dap-9 was purchased form Chiscientific, Protegrin-1 (PG-1) was purchased from the core facility at Emory university, and n-hexadecane was purchased from Fisher Scientific. All reagents were used as received with no further purification. The lipid was dissolved in n-hexadecane at a concentration of 200 mg/ml, and was used immediately. Compositions used were either 100% DOPC, or 3:1 DOPC:DOPG. Bilayers were formed over a 100 micron diameter aperture in a polystyrene cup (Warner Instruments). The chambers and the cup were washed in sodium dodecyl sulfate (SDS) and distilled water before each experiment. The cup was then dried under a stream of nitrogen. Lipid solution was applied over the area surrounding the aperture using a thin glass rod. The lipid solution was then spread evenly over the surface using a stream of nitrogen. The cup was inserted into a thermally conductive chamber (Warner Instruments) and the cup and chamber sides were both filled with 1 mL of 0.1M KCl solution, pH 7.4. Ag/AgCl electrodes were inserted in small electrode wells containing 3M KCl which were then connected to the cup and chamber liquid using salt bridges containing 3M KCl with 2.5% agar. The chamber was then placed in a BLM-TC thermocycler (Warner) and allowed to reach 37°C. Before forming a bilayer, we checked that the current was greater than 1000 pA, indicating that there was no obstruction blocking the aperture. The lipid was then painted over the aperture again using a glass rod as an applicator. The formation of a bilayer was indicated by a sharp drop in current accompanied by an increase in bilayer capacitance. Current and capacitance were measured using a BC-535 patch clamp amplifier (Warner Instruments). Bilayers were required to reach a capacitance value greater than 40 pF to be used in our experiments. Once these conditions had been met and the capacitance and current had remained stable for at least 15 minutes, a transmembrane potential of 50 mV with respect to the *cis* side was applied across the bilayer, which was then allowed to again equilibrate until the capacitance and current were stable for 15 minutes, while making a baseline recording. Finally, the peptides were added to the *cis*-side to a final concentration of 7 µM. The data was then collected from the amperimeter through a Digidat 1440 data digitizer (Axon Instruments) at a sampling rate of 150 kHz and analyzed using the Clamplex 10.0 software (Molecular Devices Corp.). Signal filtering was achieved using the 4-pole bessel filter built into the amplifier unit, at a setting of 2 kHz.

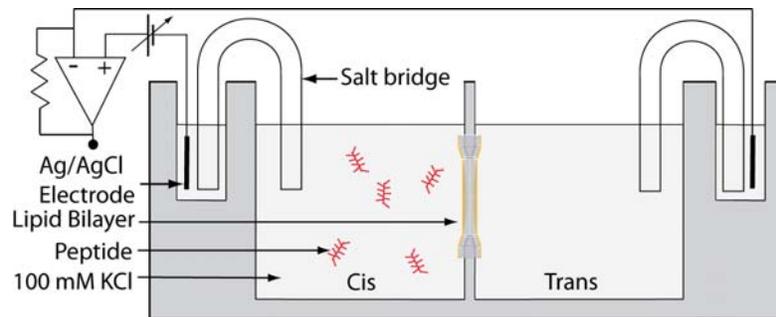

**Figure 1** Schematic representation of the planar lipid bilayer setup used to measure ionic currents across the bilayer induced by peptides added to the *cis* chamber.

*Cell isolation procedure*
Umbilical cords obtained after vaginal and cesarean delivery were placed in saline solution containing 130 mM NaCl, 4.7 mM KCl, 1.17 mM $KH_2PO_4$, 1.16 mM $MgSO_4$, 24 mM $NaHCO_3$, 2.5 mM $CaCl_2$, pH 7.4 at 4°C and immediately transported to our laboratory and stored at 4°C. The arteries were dissected from the Wharton´s jelly just before the cell isolation procedure. Human umbilical artery (HUA) smooth muscle cells were obtained following a method described by Klockner (40) and later modified in our laboratory in order to diminish the enzyme content in the dissociation medium (DM) (41). Briefly, a segment of HUA was cleaned of any residual connective tissue, cut in small strips and placed for 15 min in a DM containing 130 mM NaCl, 1.2 mM $KH_2PO_4$, 5 mM $MgCl_2$, 5 mM HEPES, 6 mM glucose, pH was adjusted to 7.4 with NaOH. The strips were then placed in DM with 2 mg/ml collagenase type I during 25 min, with gentle agitation, at 35°C. After the incubation period the strips were washed with DM and single HUA smooth muscle cells were obtained by a gentle dispersion of the treated tissue using a Pasteur pipette. The remaining tissue and the supernatant containing isolated cells were stored at room temperature (~20°C) until used.

*Patch-clamp recordings*
HUA smooth muscle cells were allowed to settle onto the coverglass bottom of a 3 ml experimental chamber. The cells were observed with a mechanically stabilized, inverted microscope (Telaval 3, Carl Zeiss, Jena, Germany) equipped with a 40X objective lens. The chamber was perfused for 15 min, at 1 ml.min$^{-1}$ by gravity, with the appropriate saline solution (see composition later) before the patch-clamp experiment was started. Application of test solutions was performed through a multibarreled pipette positioned close to the cell investigated. After each experiment on a single cell, the experimental chamber was replaced by another one containing a new sample of cells. Only well-relaxed, spindle-shaped smooth muscle cells were used for electrophysiological recordings. Data were collected within 4-6 h after cell isolation. All experiments were performed at room temperature (~20°C).

The standard tight-seal whole-cell configuration of the patch-clamp technique was used. Glass pipettes were drawn from WPI PG52165-4 glass on a two-stage vertical micropipette puller (PP-83, Narishige Scientific Instrument Laboratories, Tokyo, Japan) and pipette resistance ranged from 2 to 4 Mohms. Ionic currents were measured with an Axopatch 200A amplifier (Axon Instruments, Foster City, CA). Currents were filtered at 2 kHz, digitized (Digidata 1200 Axon Instruments, Foster City, CA) at a sample frequency of 20 kHz. The experimental recordings were stored on a computer hard disk for later analysis.

*Solutions for whole cell experiments*: The extracellular saline solution used for recording whole-cell ionic currents contained 130 mM NaCl, 4.7 mM KCl, 2.5 mM $CaCl_2$, 6 mM glucose, and 5 mM HEPES. The pH was adjusted to 7.4 pH using KOP.
The composition of the intracellular pipette solution contained 130 mM KCl, 5 mM $Na_2ATP$, 1 mM $MgCl_2$, 0.1 mM EGTA, and 5 mM HEPES. The pH was adjusted to 7.3 with NaOH.
*Solutions for cell-attached experiments*: The tip of the peptide was filled by capillarity with a solution composed of 140 mM KCl, 0.5 mM $MgCl_2$, 10 mM HEPES, 10 mM Glucose, and 1 mM $CaCl_2$. The pH 7.4 was adjusted using KOH. The rest of the pipette is filled with the same solution plus Arg-9 peptides at a concentration of 7 µM. The bath solution was composed of 140 mM KCl, 0.5 mM $MgCl_2$, 10 mM HEPES, 10 mM Glucose, and 1 mM EGTA. The pH was adjusted to 7.4 using KOH.
*Solutions for inside-out experiments*: Both bath and pipette solutions were the same composed of 140 mM KCl, 0.5 mM $MgCl_2$, 10 mM HEPES, 10 mM Glucose, 1 mM EGTA. The pH 7.4 was adjusted using KOH.

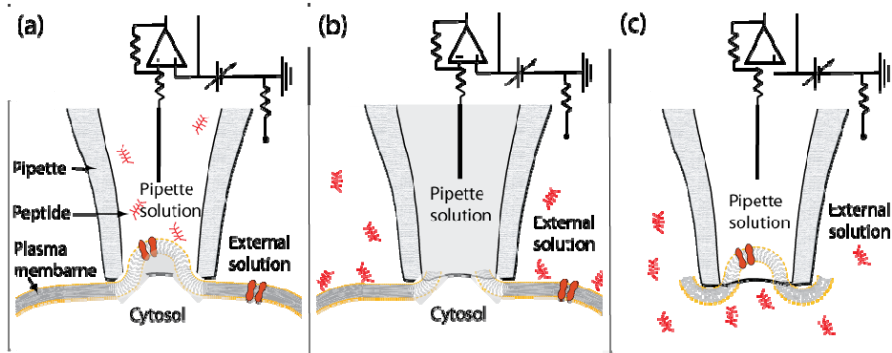

**Figure 2** Schematic representations of the patch clamp configurations used to measure ionic currents across mammalian cell membranes induced by the Arg-9 peptides: (a) *Cell attached* (b) *Whole-cell* and (c) *Inside out*.

# Results and discussions

## Theoretical Model

Recently, we proposed a mechanism for the translocation of the HIV-1 Tat peptide based on molecular dynamics simulations. This mechanism shows how these peptides are able to passively diffuse across phospholipid bilayers (20). The translocation mechanism of these peptides can be described as composed of four basic steps: (a) The peptides bind to the surface of the bilayer, attracted by the phosphate groups of the phospholipids. (b) As the surface concentration of peptides increases, the arrangement of lipids is strongly distorted compared to that in the resting membrane. (c) An arginine side chain translocates to the distal layer nucleating the formation of a water pore. (d) A few peptides translocate by diffusing on the surface of the pore and the pore closes. These simulations highlight the central role of the arginine amino acids, such as their strong attachment to the phosphate groups and their ability to induce strong distortions to the structure of the phospholipid bilayer. It has been reported that Arg-9 peptides internalize more efficiently than Tat peptides. Therefore, a natural question is if peptides composed entirely of arginine amino acids such as Arg-9 would follow a similar translocation pathway as the Tat peptide.

We performed a molecular dynamics simulation of a system composed of 4 Arg-9 peptides, 92 1,2-di-oleoyl-sn-glycero-3-phosphocholine (DOPC) phospholipid molecules and 8795 water molecules. The simulations were started from equilibrated lipid bilayer systems and extended for 500 ns. In Fig. 3 are shown four snapshots extracted from this MD simulation in which it can be seen that Arg-9 follows the same mechanism as was previously described for Tat.

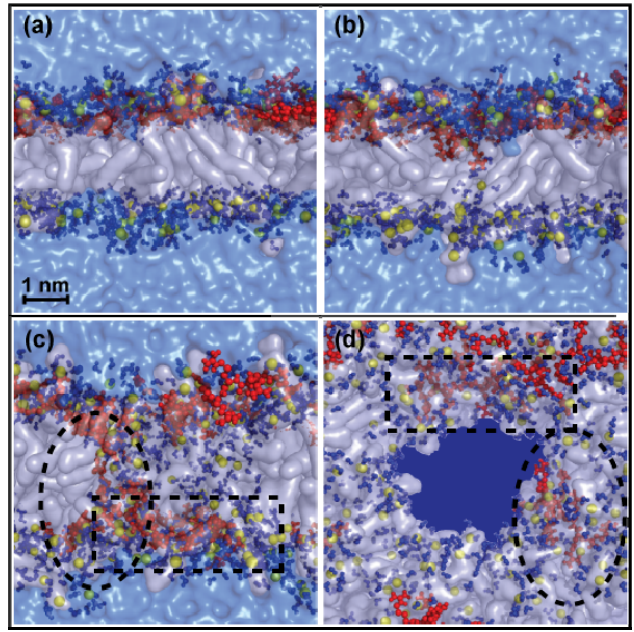

**Figure 3** Four snapshots of a molecular dynamics simulation (a) A lateral view that shows how the peptides are bound to the membrane before translocation. (b) The translocation of an arginine amino acid surrounded by water molecules that nucleates the formation of a water pore. (c) Lateral view of the pore, translocated peptide surrounded by a dotted square line and a translocating peptide circled by an oval dotted line. (d) Top view of the pore. The simulated system consists of 4 Arg-9 peptides, 92 DOPC

phospholipids and 8795 water molecules. The phospholipid molecules are represented with transparent white surfaces, the phosphate atoms are in yellow spheres, the peptide molecules are in red, any water molecule at a distance of less than 3.5 Å from any phospholipids or amino acid atom is colored in solid blue, and the rest of the water molecules appear as a transparent blue surface.

During the first 50 ns the Arg-9 peptides bind to the phosphate groups in the proximal layer of the lipid bilayer. The guanidinum groups of the arginine side-chains make hydrogen bond contacts with various oxygen atoms of the lipid headgroups and the peptide is partially immersed into the bilayer, at the level of the glycerol groups. After 100 ns an arginine side-chain is attracted to a phosphate group in the distal layer of the bilayer. This phosphate group and the Arg side chain carry some solvation water molecules with them, so they are never fully desolvated. During this period of time a water chain crossing the bilayer is formed and a toroidal pore is nucleated. One of the Arg-9 peptides translocates by diffusing on the surface of the pore. In this calculation the toroidal pore remains open toward the end of 500 ns. We expect the complete translocation and closing of the pore to occur on the microsecond time scale. The radius of the toroidal pore varies in size having a maximum diameter of 2.5 nm. The pore surface is lined up with phosphate headgroups. The mechanism of arginine side chain insertion and pore formation is identical to the mechanism found for the Tat peptide (20).

The destabilization of lipid membranes and the formation of pores induced by these peptides should lead to an increase in the ionic permeability of phospholipid bilayers and the cell membrane. To validate this prediction of the model, we investigated experimentally the Arg-9 permeabilizing effects in lipid membranes and mammalian cells using two electrophysiological setups as described below.

## Permeabilization of model lipid membranes

If CPPs are able to destabilize phospholipid bilayers producing transient pores across them, then ionic permeabilization should be observed. To test this hypothesis on model lipid membranes we made phospholipid bilayers using the *planar lipid bilayer* (also known as the *black lipid membrane*) method (29). The basic idea of this setup (more details are described in the method section) is that the bilayer is made across a 100 μm diameter hole that separates two chambers filled with a saline solution. In each chamber an electrode is introduced to generate an electrostatic potential across the membrane and measure ionic currents. The chamber where the active electrode is introduced is conventionally called the *cis* chamber and the other chamber is called the *trans* chamber. The *cis* chamber is held at 50 mV relative to the *trans* chamber. The peptides are added to the *cis* chamber because according to the model it is expected that they would be more likely to cross the bilayer from the higher to the lower electrostatic potential, which corresponds to the potential difference between the exterior and interior of the cell.

Addition of micromolar concentrations of Arg-9 to the aqueous solution bathing a planar bilayer membrane led to an increase of membrane conductance in all experiments (n > 50, where n is the number of experiments). Fig.4 (a) shows the permeabilization of a phospholipid bilayer composed of a lipid mixture of DOPC:DOPG (3:1) after the addition of Arg-9 at a concentration of 7μM to the *cis* chamber. As seen in Fig. 4 (a) after the peptide is added to the *cis* chamber the ionic permeabilization across the membrane increases. A similar effect is shown in Fig. 4 (c) for bilayers entirely composed of DOPC phospholipid molecules. This proves that anionic phospholipids such as DOPG are not strictly required for peptide permeabilization of the bilayer since the peptides are still able to interact and permeabilize purely zwitterionic DOPC phospholipid bilayers. The permeabilization increases both continuously and with discrete jumps. The recordings show that the permeability increases and reduces several times until finally the membrane breaks permanently. The transient current spikes were quite abrupt and most of the time did not resemble recordings of well-defined ionic channels such as voltage-dependent ion channels.

Positive ions such as divalent $Ca^{2+}$ ions would be expected to displace the arginine amino acids or screen the underlying membrane phosphate groups. The permeability of the plasma membrane produced by the peptides should therefore be reduced upon the addition of calcium chloride. Indeed, as shown in Fig. 4 (b), when calcium chloride was added to the solution, the permeabilization of the membrane reduced to a low baseline level. Moreover, as shown in Fig. 4 (d), when the peptides were added after adding 100 mM of $CaCl_2$ to the solution, there was no increase in the permeability of the bilayer, consistent with the ability of the $Ca^{2+}$ ions to mask the phosphate groups from the peptides. This finding highlights the importance of electrostatic interactions in the initiation of pore formation.

The MD simulations suggest that reducing the length of the amino acid side chain and replacing the guanidinium group with an amine group would diminish the efficiency of peptide translocation. Therefore, we tested the effect of Dap-9 (composed of nine 2,3-diaminopropanoic acid residues, Fig. 5), which has the same charge as Arg-9 but much shorter side chains and amine groups instead of guanidinium groups. As seen in Fig. 4 (e) Dap-9 does not permeabilize the phospholipid membrane.

We also performed a negative control shown in Fig. 4 (f), using the same solution but with no peptide addition and no current increase was observed.

Fig. 6 (a) shows the current induced by Arg-9 across the bilayer for a system with the same composition and held at the same voltage as in Fig. 4 (a). Fig. 6 (b) and (c) show the dependence of the current on the voltage. Each interval characterizes a common state observed during the permeabilization of the membrane: control I-V represents a control measurement before the peptide is added; I-V (i) is a measurement after the addition of the peptide but before the steady increase of the conductance across the bilayer; and I-V (ii) is a measurement during the steady increase of conductance across the lipid bilayer. As seen in the figure, current fluctuations were seen initially after the peptides were added, but the baseline current remained unchanged and there was no significant difference in the average current versus voltage recorded (Fig. 6 (b) I-V (i)), compared to the same measurement before adding the peptides (Fig. 6 (b) I-V control). However, after a few minutes the permeabilization became permanent (the current base line increased) and there was a marked increase in slope of the current versus voltage for I-V (ii) relative to that for control I-V. In Fig. 6 (a) and (c) it can be seen that after adding 100mM of $CaCl_2$ the permeability decreased. Furthermore, after waiting for more than

100 min the permeability dropped to a very low baseline values as can be seen in Fig. 6 (c) I-V (iv).

These results could be interpreted in the following way: After the peptides are added to the solution they start binding the bilayer creating isolated local transient pores across the bilayer. Consequently, the current base line initially remains at the same level as that before adding Arg-9. The slope of the current vs. voltage plot in I-V (i) remains almost unchanged compared to that for the I-V control, indicating that the permeability increases transiently while the baseline remains most of the time unchanged. After a few minutes the surface density of membrane-bound peptides increases, destabilizing the membrane more strongly and more permanently as is reflected by the increase in the base line current at I-V (ii) and by the increase in the slope of the current versus voltage. After the addition of $CaCl_2$ the permeability decreases indicating that $Ca^{2+}$ ions might be competing for phosphate groups and displacing the Arg-9 peptides from the surface of the membrane. This results in the resealing of the phospholipid bilayer.

In the inset pictures of Fig. 6 and in Fig. 7 are shown amplified current traces. One can see discrete current jumps that resemble the behavior of ion channels in cells and rapidly fluctuating current jumps that can be clearly distinguished from most ionic channels. These latter types of signals can be used as a foot print to recognize the peptide interaction with the cell membrane besides the expected steady permeabilization increment with time.

Our measurements show that most of the time the current fluctuates rapidly, while in some cases more stable current jumps that resemble ion channels can be observed. If we assume that these jumps are associated with the formation of stable pores across the membrane, we can estimate the average radius of these pores by measuring the conductance. To characterize this behavior we show in Fig. 2 of the supplementary material a histogram of the amplitude of those current jumps over 118 events. Assuming that the pore conductivity is equal to that of 100mM of KCl, we can calculate the pore radius for a conductance G of 4.6 pA/50 mV=0.92 $10^{-10}$ S using $\sigma=Gh/A$. Where $\sigma$ is the conductivity (~1.6 S $m^{-1}$); A is the area of the pore; G is the conductance (~0.92 $10^{-10}$ S); and h is the membrane thickness (~6 nm). The estimated pore diameter for this current value is d~ 2 $\sqrt{(Gh/\sigma\pi)}$=0.66 nm, which compares well with the pore diameter observed in the simulations.

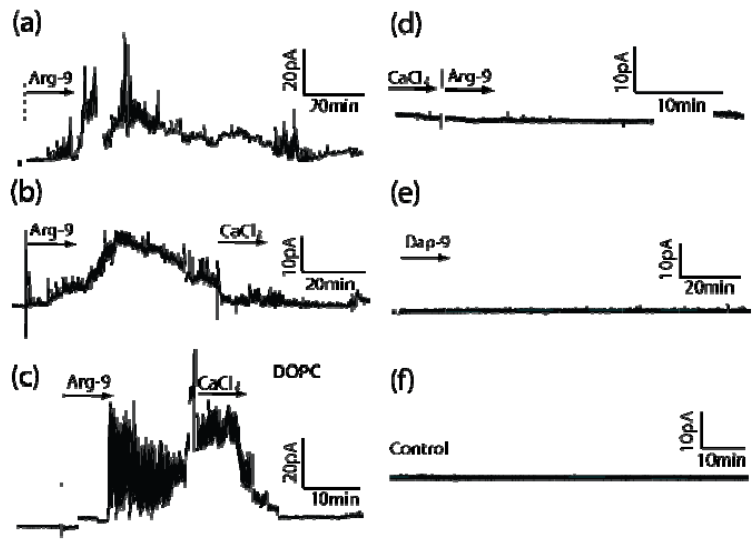

**Figure 4** Permeabilization of phospholipid bilayers composed of a lipid mixture of DOPC:DOPG (3:1) (a) and (b) after the addition of 7μM of Arg-9 to the *cis* chamber, (c) phospholipid bilayer composed entirely of DOPC, (d) lipid mixture of DOPC:DOPG, the CaCl$_2$ ions are added to the solution before the addition of the peptide, (e) control measurement on a lipid mixture of DOPC:DOPG adding the Dap-9 peptide, (f) control on a lipid mixture of DOPC:DOPG measurement without peptides. The arrow's origin indicates the time at which the peptides or the CaCl$_2$ are added to the solution. The potential of the *cis* chamber relative to the *trans* chamber (the holding potential) is 50mV. The ionic concentration is 100mM of KCl and the pH is 7.4.

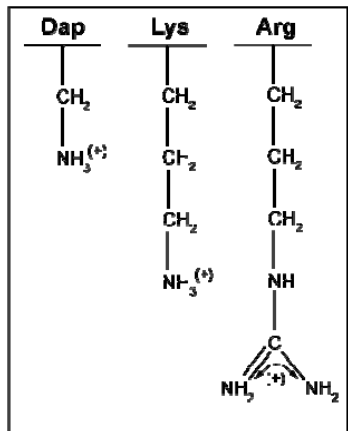

**Figure 5** Side chain structures of the α-amino acids Dap, Lys, and Arg.

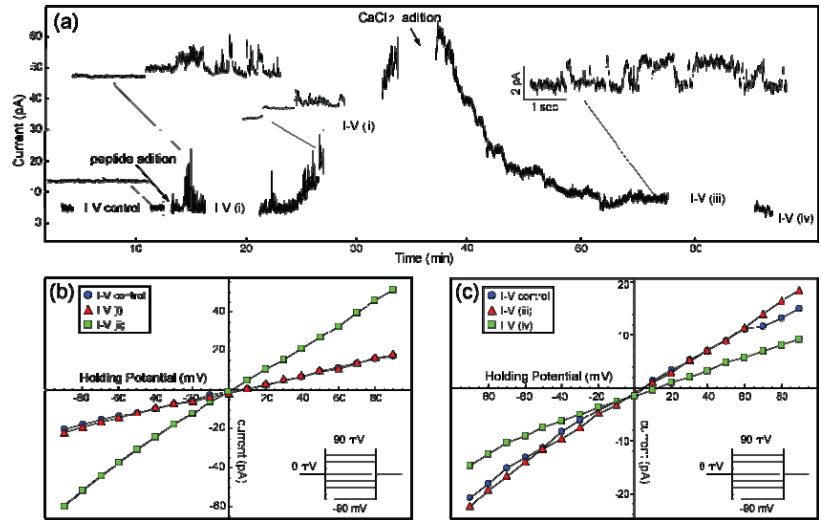

**Figure 6** Current across a planar lipid bilayer under the same conditions as in Fig. 4 (c, d). (a) An average over ten measurements of current as a function of voltage were carried out at six different time points: (I-V control) before adding the peptides, (I-V i) 10 minutes after adding the peptide, (I-V ii) 40 minutes after adding the peptide, (I-V iii) 30 minutes after adding $CaCl_2$, and (I-V iv) 70 minutes after adding $CaCl_2$. The inset pictures show amplified sections of the current trace, the scaled time fraction if enclosed by a dotted square, when time and current are both scaled the scale is included in the inset picture. (b) and (c) Average current as a function of voltage. The voltage is incremented by 10mV between -90mV to 90mv and the potential is held for 150ms at each voltage. The points in (b) and in (c) are the average over 10 consecutive measurements taken at each of the intervals indicated in (a). It can be seen that after the peptide is added there are currents jumps before (I-V i) but there is not much difference in average between I-V control and (I-V i). However, after the permeabilization becomes permanent and the base line current starts to increase there is a marked increase in the slope of the current versus voltage (I-V ii) relative to the control I-V (I-V control). After addition of $CaCl_2$ the ionic permeabilization reduces. In the last I-V record (I-V iv), the current is even lower than the control current. This indicates that $CaCl_2$ reduces the permeability of the membrane to essentially zero.

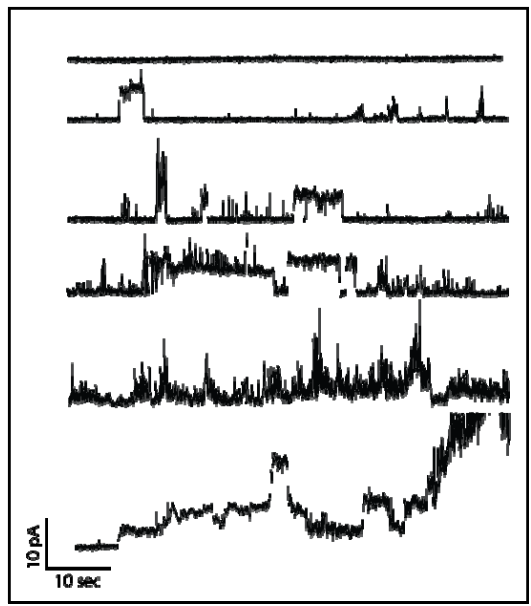

**Figure 7** Fractions of current traces after the addition of 7μM of Arg-9 to the *cis* chamber as the time progress. The planar lipid bilayer is composed of a lipid mixture of DOPC:DOPG (3:1). It can be seen that as the time increases the permeabilization has a noisier trend.

The change in permeability induced by polyarginine peptides is indicative of the relevance of arginine amino acids in the local destabilization of the bilayer and pore formation capabilities of CPPs. Although Arg-9 is the focus of this experimental work, we are currently investigating the effects of Tat, penetratin, and PG-1. The first is also a highly hydrophilic CPP, the second is a slightly more amphiphilic CPP, and the third is a pore forming arginine-rich amphiphilic antimicrobial peptide. We also observed in all these cases an increase in the permeabilization of the phospholipid bilayers upon addition to a final concentration of 7 μM (Fig. 1 of the supplementary material). These experiments are still underway.

## Permeability across mammalian cells

Our computer simulations allowed us to develop a hypothetical model and unveil some microscopic details. However the simulated time that can be achieved and the size of the systems is small compared to the experiments on actual phospholipid bilayers. The increase in precision obtained with actual systems, although critical to verify any hypothesis, usually comes at the cost of losing atomistic details. In this way each system represents a different level of detail that complements the others. We have shown that the translocation mechanism is consistent with the permeability changes measured on lipid bilayers. Investigating the effect of these peptides on the cell is the final level of complexity that we chose to explore. The planar phospholipid bilayers systems do not contain the complexity and machinery present in cells. However, we can still establish a cause and effect correlation between the addition of CPPs to a cell, under various configurations, and the permeability behavior of the cells.

The results obtained in lipid bilayers, where it is possible to see that current is able to flow across the membrane in the presence of the peptide, encouraged us to test the peptide effects on cell membranes of human smooth muscle cells and osteosarcoma cells, although these systems are structurally and functionally much more complex.
To accomplish this task we used the patch-clamp technique, in different configurations (whole-cell, cell-attached and inside-out, each of them described in the methods section) in order to measure ionic currents induced by the peptide on the cell membrane.

*Whole cell:* **(global permeabilization of the cell)**

We studied the effects of Arg-9 on isolated human umbilical artery (HUA) smooth muscle cells in the whole-cell configuration (see methods for details). In this case the patch pipette containing a saline solution of a composition similar to the intracellular medium was tightly sealed to the membrane; after a soft suction the membrane patch under the pipette tip was ruptured and the solution comes in contact with the intracellular medium. An electrode placed inside the pipette measured the current across the whole plasma membrane of the cell and, at the same time, was able to control the voltage across the membrane. In this case the cells were clamped at a holding potential of -50 mV, hence evoking a macroscopic holding current, which we measured before and after adding Arg-9 to a final concentration of 0.07 or 7 µM to the bath solution. Fig. 8 (a) shows a typical time course of the current at -50 mV where we see that the peptide slowly increased the magnitude of this inward current. The effects of both concentrations of Arg-9 were variable among the cells tested, and some of them did not show any significant changes (5 out of 7 cells tested for 0.07 µM; and 2 out of 9 cells tested for 7 µM). In the cells where the peptide produced an effect, the average holding current showed a significant increase over the control current measured in the absence of the peptide. The mean current increase induced by 7 µM concentration of Arg-9, expressed as per cent of the control current, is $126 \pm 45$ % (n=7). We also observed that Arg-9 induced instantaneous current jumps like the ones shown in Fig. 8 (b). A similar behavior could also be observed at other membrane potentials (data not shown).

It is known that acidification of the extracellular medium could destabilize phospholipid bilayers, thereby enhancing the destabilizing effects induced by the peptide on the plasma membrane. Therefore, we tested the peptide effect on the holding current at -50 mV in an acid bath solution (pH = 5.5). Under this condition we tested 9 HUA smooth muscle cells, and in 6 of them the current increased to a much greater extent than at the physiological pH of 7.4. The mean current increase (Fig. 8 (c) and (d)) was in this case $460 \pm 130$ % (averaged over 6 experiments).

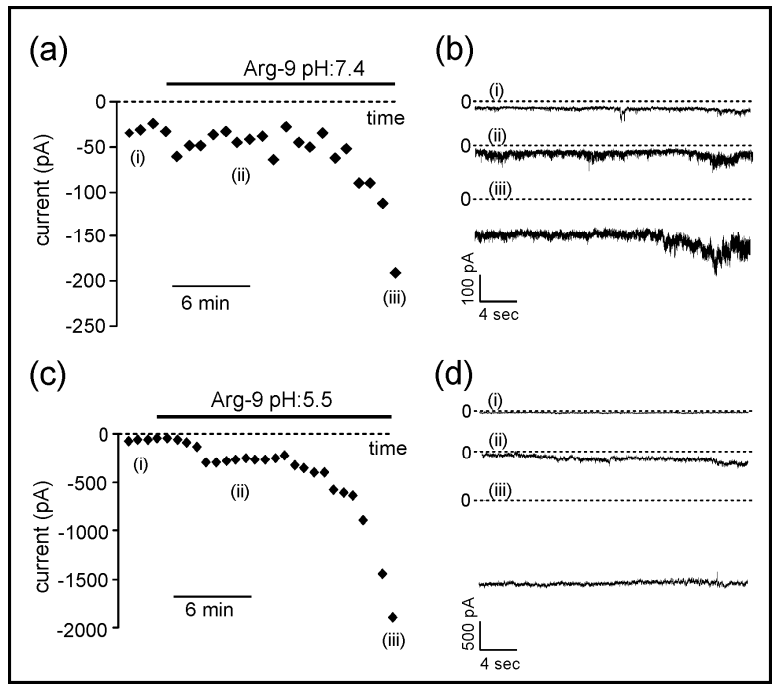

**Figure 8** (a) Time course of the increase of the inward membrane current after the addition of 7 µM of Arg-9 to the extracellular solution in the whole-cell configuration in the case of HUA smooth muscle cells. (a) Three typical recordings of this current (pH 7.4), corresponding to the points (i), (ii) and (iii) indicated in (a) and (c). If the pH of the bath solution is lowered to 5.5, the increase in the inward current is more prominent. (d) Three typical recordings of this current, corresponding to the points (i), (ii) and (iii) indicated in (c).

**Cell attached and inside out: Local cell permeabilization**

We next tested the effect of Arg-9 on cells in the cell-attached and inside-out configurations. These configurations allowed us to measure the currents induced by the peptide in membrane areas which are small compared to the whole surface of the cell.

In the *cell-attached* configuration the tip of the pipette was sealed to the cell membrane without rupturing it. Therefore, the pipette-solution was in contact with the extracellular face of the cell-membrane patch, and was maintained in this condition during the recording, while the intracellular cell content remained intact. The tip of the pipette was filled with a pipette solution without peptide, while the rest was back-filled with the same solution containing the peptide. This produces a time delay (a few minutes) before the peptide slowly diffuses into the tip and comes into contact with the membrane. The membrane potential of the cell was kept near 0 mV by bathing them with a high $K^+$ extracellular solution, but the presence of the seal allowed the application of a transmembrane voltage of 50 mV (relative to the interior of the cell) across the patch under the pipette, which was maintained constant during all the recoding time. In Fig. 9 we can observe several recordings of the current evoked by a concentration of 7 μM of Arg-9 in the pipette at different times after the seal was obtained in HUA smooth muscle cells. A concentration of 0.07 μM of the peptide was also tested (data not shown).

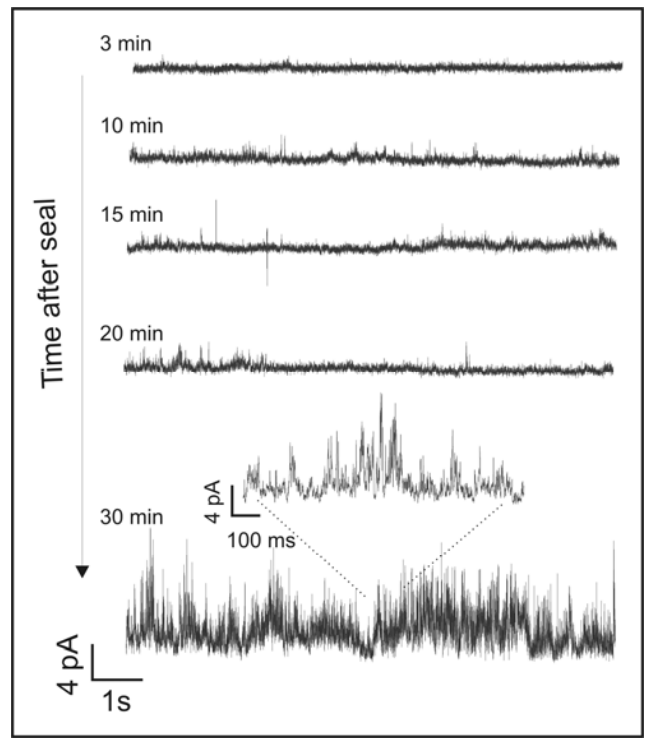

**Figure 9** Typical cell-attached recordings of the current evoked by the peptide placed inside the pipette in HUA smooth muscle cells at different times after obtaining the seal.

In the majority of tested HUA smooth muscle cells, after periods of time ranging from 10 to 35 minutes, we began to observe rapidly fluctuating current jumps that resemble the ones we previously observed in artificial phospholipid bilayers. These signals do not follow a repetitive pattern and are difficult to characterize by the typical parameters used to study ionic channels, such as mean current amplitude, open probability or dwell times. Instead, they resemble the current patterns observed on perturbed membranes of different kind of cells exposed to mechanical stress (42, 43) or in cardiac cells stimulated by photosensitizer-generated reactive oxygen species (44). This characteristic signal can be observed in 6 out of 7 cells obtained from different samples of umbilical cords, although the effect of the peptide is variable in its magnitude. We can also observe similar rapidly fluctuating current jumps in some of our control cells (without peptide in the pipette), however, with a much lower frequency of appearance than in the presence of Arg-9. Cell-attached experiments with the peptide (0.07 µM) were also made in UMR106 osteosarcoma cells with similar results.

Figure 10 shows current snapshots that compare the evolution of current with time for osteosarcoma, HUA smooth muscle cells, and phospholipid bilayers. It can be seen that in every case the ionic currents observed demonstrate a time-increasing rapidly fluctuating behavior. In the case of the phospholipid bilayer, a stepwise current can also be seen, which resembles the opening behavior of ionic channels. However, if the peptides produce discrete jumps in the cells these would not be easily distinguished from those belonging to the ion channels already present in the cells. Hence, in order to minimize the activation of voltage-operated channels, we tested the effects of the peptide using a -50 mV membrane potential, at which in these kind of cells most of these voltage-operated ionic channels present a low open probability. Testing the peptide on a biological system, such as a cell, introduces some of the variables present in complex systems, such as the presence of different kinds of ionic channels and lipid composition, that are not present in the planar lipid bilayer experiments.

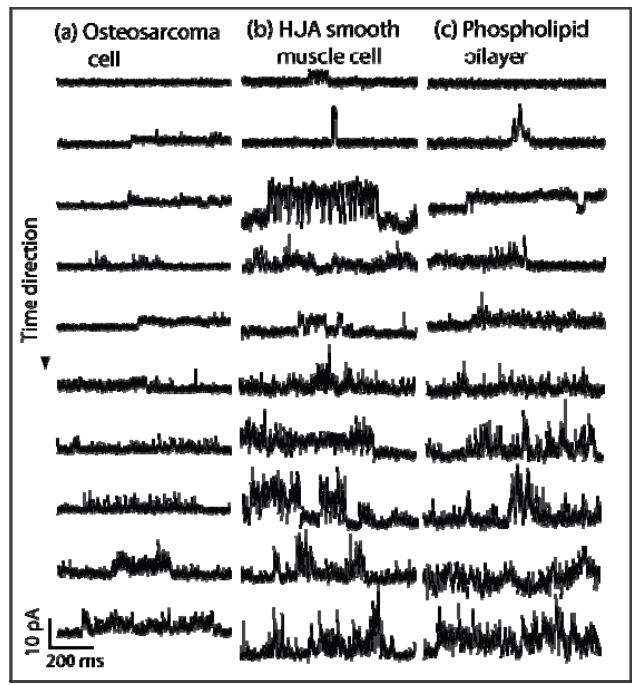

**Figure 10** Snapshots of typical recordings of the current evoked by the peptide (placed inside the pipette) at different times after obtaining the seal in the cell-attached configuration in a (a) osteosarcoma cell, (b) HUA smooth muscle cell, and a (c) phospholipid bilayer. It can be seen

that as time increases the membrane is being increasingly permeabilized and the signal reflects a rapidly fluctuating current in every case.

In the *inside out* configuration a micrometer portion of the cell surface was covered by the tip of the glass pipette, then was pulled out and only a small piece of the membrane remained in contact with the pipette. In this case the internal surface of the cell membrane was exposed to the bath solution. EGTA (1 mM), a calcium chelator was present to inhibit the activity of calcium dependent ionic channels. As in the case of the cell-attached configuration, the recordings were obtained in a symmetric high-potassium saline solution. When a transmembrane potential of -50 mV (outside positive) was applied, a stable holding current appeared. Immediately after the addition of Arg-9 to a concentration of 7 µM to the bath solution (now in contact with the intracellular side of the membrane), the negative net current measured at -50 mV significantly changed reaching more positive current values (Δcurrent = 75±18 pA n=13). After that, we observed again the occurrence of rapidly fluctuating, unstable, variable current jumps, such as those observed in the last part of the recording shown in Fig 11. These two kinds of behaviors were seen only in the presence of Arg-9, and were similar in all 13 cells tested. They were not observed in control inside-out patches. The peptide effect seems to have two phases in this patch clamp configuration: an immediate improvement of seal resistance followed by an increase of the patch noise, reflecting the destabilizing effect observed in the other configurations. This could be partially explained by the observations made by Chico et al. (45) who reported that significant amounts of poly-D-Arg are retained by plastic and glass surfaces. This finding could explain our results, because Arg-9 would be able to improve the seal by increasing the interaction between the cell membrane and the glass pipette, which is the base of the establishment of a high resistance seal. This effect does not exclude the fact that Arg-9 could then penetrate, by destabilization, the cell-membrane, as could be inferred by the occurrence of the typical noisy current but the real concentration of the effective peptide action could be much less than the 7 µM peptide tested concentration.

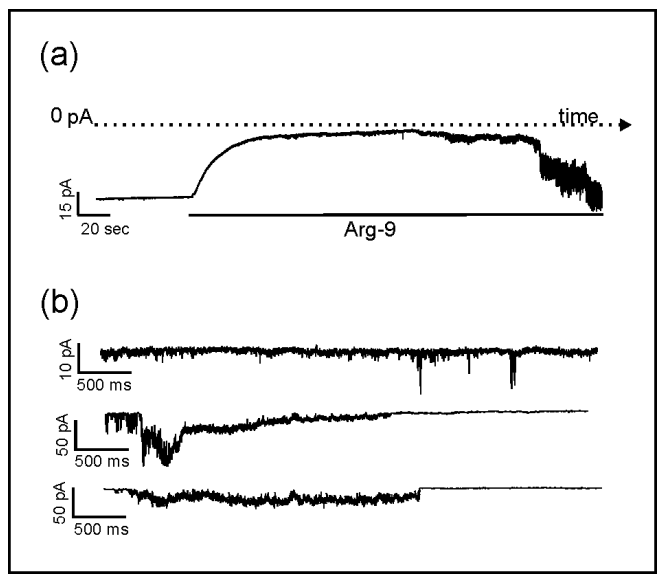

**Figure 11** Typical recording of Arg-9 effects in an inside-out patch obtained from a HUA smooth muscle cell in symmetrical ionic conditions. (a) recording showing the current measured at 50 mV (outside negative) and its decrease induced by the Arg-9 addition at a concentration of 7 μM. After a few minutes, the current trace becomes rapidly fluctuating in presence of the peptide. The variability of these current jumps amplitude can be observed in the three traces of panel (b) extracted at different time intervals.

## Conclusions

In this work, we show experimentally that Arg-9 peptides destabilize and permeabilize planar phospholipid bilayers and cell membranes consistent with our theoretical model.

The proposed translocation mechanism for arginine rich cell-penetrating peptides involves the destabilization and the formation of transient pores in model lipid bilayers. This theoretical mechanism highlights the central role of the arginine amino acids for the peptide binding to the phospholipids and the nucleation of transient pores across phospholipid bilayers. However, it is usually assumed that to interact and permeabilize phospholipid membranes the peptides must have a high degree of hydrophobicity such as is the case of arginine-rich antimicrobial peptides. Nevertheless, we show that CPPs composed entirely of arginine amino acids, despite being charged and highly hydrophilic, are capable of interacting, destabilizing and permeabilizing phospholipid bilayers and mammalian cell membranes.

We performed a molecular dynamics simulation of Arg-9 that reproduces the main steps for translocation previously observed for the Tat peptide (20). The peptides bind to the surface of the bilayer, attracted by the phosphate groups of the phospholipids. These interactions produce large local distortions to the phospholipid bilayer compared to its resting structure. These distortions reduce the hydrophobic free energy barrier of the bilayer and an arginine side chain translocates attracted by phosphate groups on the distal side. This helps the nucleation and the formation of a toroidal pore. Once the pore is formed, the peptides translocate by diffusing on the surface of the pore. The simulations highlight the central role of the arginine residues. This is in accordance with experimental evidence presented by Mitchell et al. (46), who showed that six or more guanidinium groups are required to efficiently penetrate the cell membranes of different kinds of cells. Moreover, they reported that Arg-9 peptides internalize more efficiently than Tat peptides, probably due to the higher content of arginine amino acids. Despite intense research, details of the cell penetration mechanism of poly-arginine peptides are still under discussion. Futaki and coworkers have recently reviewed the subject (47) and concluded that proteoglycans may play a crucial role as primary receptors for peptide internalization, and that both endocytosis and direct membrane translocation are possible entry routs for arginine rich peptides.

Still, some of the most important points that seem to divide opinions are whether the process is dependent on cell metabolic energy, and whether it requires vesicle formation (i.e., endocytosis (48) and macropinocytosis (49), among others). There is experimental evidence showing the inhibition of poly-arginine uptake obtained by blocking the cell metabolism with sodium azide (46), by lowering the temperature to 4 ºC (50), and by macropinocytosis and endocytosis inhibitors (12, 49, 51). However, there are other reports that suggest that cell penetration is independent of metabolic cell energy. Confocal microscopy experiments showed that a significant amount of diffuse peptide internalization could also be observed, probably due to direct peptide translocation. Furthermore, it has been recently shown, with genetically engineered cells, that knocking down clathrin-mediated and caveolin-mediated endocytosis (52) does not affect the ability of Tat to enter cells. In this case, there is an open discussion on a broad spectrum of possible mechanisms (4, 50, 53). The results presented here can explain the observations of direct translocation of the peptides through a mechanism that is independent of metabolic cell energy.

Our theoretical observations show that Arg-9 is able to nucleate transient pores that would allow ions to flow across it. Therefore, applying an electrostatic potential across the bilayer should induce a net ionic current that could be detected experimentally. We confirmed this prediction by measuring currents across planar phospholipd bilayers and both cultured and freshly isolated mammalian cells. The currents induced by Arg-9 across planar phospholipid bilayers are most of the time rapidly fluctuating and in some cases they present discrete jumps that resemble the typical behavior of ion channels. The observed pattern of ionic currents induced by Arg-9 is quite similar in the three tested systems indicating that electrophysiological methods are useful to study the effect of these peptides on membranes. Future studies will be directed to the electrophysiological characterization of this mechanism. Patch clamp and planar lipid bilayers have been useful to characterize several peptides that permeabilize cell membranes such as, puroindolines (54), alamethicin (55, 56) and cobra cardiotoxin (57), among others.

The electrostatic attraction between the arginine amino acids and the phosphate groups of the membrane is a pivotal step of the translocation model. Therefore, different ions and counter ions can have a strong effect on the translocation of these peptides if they are able to screen these interactions. If the $CaCl_2$ concentration is increased (100 mM) after adding the peptides, the permeabilization initiated by Arg-9 is strongly reduced. Moreover, no permeabilization of the phospholipid bilayer is observed if Arg-9 is added to a solution with a high $CaCl_2$ concentration. We speculate that calcium ions may be competing for the phosphate groups reducing in this way the binding of the peptides to the surface of the membrane.

The length of the arginine side chain facilitates the insertion of the amino acid chains. For shorter amino acid side chains, like 2,3-diaminopropanoic acid (Dap), the translocation of one side chain towards phosphates on the distal layer would require either a much thinner membrane or the translocation of more side chains at the same time, making this statistically much less favorable. Previous experimental reports show that by increasing the length of the side chain, the translocation of peptides increases significantly (28). Furthermore, if the guanidinium group is replaced by an amine group the uptake is in this case reduced (46). Therefore, we measured the effect on the current produced by Dap-9 which has the same charge as Arg-9, but much shorter side chains and amine groups instead of the guanidinum groups. No permeabilization of the phospholipid bilayer was observed for this peptide.

The results obtained using the patch-clamp technique in the whole-cell configuration on human vascular smooth muscle cells confirm that Arg-9 is able to permeabilize the cell membrane. This effect is stronger under acidic conditions (pH=5.5).

Moreover, the results obtained with two other patch clamp configurations (cell-attached and inside-out patches) suggest that Arg-9 perturbs the membrane patch, thus producing a variable and rapidly fluctuating current that resembles the interaction between the peptide and planar lipid membranes. Interestingly, although an artificial phospholipid bilayer is a very simplified model of the cell-membrane, the measured currents share some common patterns. These observed currents might be due to the disorganization of the phospholipid bilayer caused by the interactions between positive charges of the peptides and the negative phosphate groups of the phospholipids. In the cell, if we consider the complete structure of the cell membrane, the peptides could also have a strong affinity for other negative membrane components like glycolipids or protein bound entities. Several authors have pointed out the importance of these interactions for the efficacy of the peptide internalization (14, 58-60). Therefore, to unveil the complexity of the CPPs translocation mechanism it is crucial to put together data from different theoretical and experimental approaches to allow a clear identification of each phase of the mechanism or mechanisms of translocation. The patch-clamp technique seems to be a very good technical choice to study the effect of these peptides mostly at the step of its interaction with the plasma membrane.

The patch-clamp experiments are performed under saline conditions near the physiological ones, in the presence of $Ca^{2+}$ at a 2.5 mM concentration, where it is possible to see the effects of the peptides. However, the inhibitory effect of $Ca^{2+}$ ions, observed on planar lipid bilayers, could modify the influence of the Arg-9 peptides. It would be interesting to characterize the effect of these ions on cells.

The physiological relevance of poly-arginine peptides on cell membranes and its ability to penetrate this selective barrier is interesting. However, these molecules have important applications beyond their capabilities as carriers. The poly-arginine peptide could have itself therapeutic use. For example, Uemura et al. (61, 62) have reported recently that heptamers of arginine inhibit vascular smooth muscle proliferation and reduced myointimal hyperplasia, a vascular response to injury involved in different pathological conditions like restenosis and atherosclerosis. Their findings open new opportunities where poly-arginine peptides could be an important tool to improve the treatments of vascular disease.

We have demonstrated using theoretical and experimental results that a peptide made up entirely of arginine residues, a highly hydrophilic and charged molecule, is able to interact and permeabilize biological membranes. Furthermore, we have shown that the length of the side chains, the guanidinium group and counterions present can have a profound effect on the ability of these peptides to cross the membrane. Understanding the mechanism by which cell penetrating peptides disrupt the plasma membrane, will contribute to the fundamental knowledge of internalization mechanisms of charged molecules through lipid membranes, and the design of more efficient cargo carrying molecules which could lead to better drug delivery candidates.

**Acknowledgments**: AEG acknowledges support from National Science Foundation Grant DMR-0117792 and from Rensselaer Polytechnic Institute. RSK acknowledges support from the NSF NIRT program (CBET 0608978). JL acknowledges support from an NIGMS Biomolecular Science and Engineering training program fellowship.
The authors acknowledge L. Rimorini, R. Roldan Palomo, S. Saleme for patch clamp technical assistance and discussions; P. Urdampilleta for the collection of the umbilical cords; Dr. M. S. Molinuevo for the UMR106 osteosarcoma cells.